\documentclass{PoS}

\title{Understanding the structure of the proton: From HERA and Tevatron to LHC}

\ShortTitle{Understanding the structure of the proton: From HERA and Tevatron to LHC}

\author{\speaker{Laurent SCHOEFFEL}%
         \thanks{A footnote may follow.}\\
        CEA Saclay, Irfu/SPP, France \\
        E-mail: \email{laurent.schoeffel@cea.fr}}

\abstract{
Understanding the fundamental structure of matter requires an understanding of how
quarks and gluons are assembled to form hadrons and of the structure of the
protons which are the colliding particles at LHC.
The arrangement of quarks and gluons
inside nucleons can be probed by accelerating electrons, hadrons or nuclei to precisely 
controlled energies, smashing them into a target nucleus
and examining in detail the final products.
The LHC physics program is rich and has been widely
described. It encompasses the searches for new
particles up to masses of several TeV, including the elucidation of
electroweak symmetry breaking and the possible observation of new
symmetries at higher scales, and precision measurements of fundamental
parameters in the electroweak and strong gauge sectors.
Obviously, this program requires a precise understanding 
of the structure of the proton in terms of
quarks and gluons, obtained from HERA and Tevatron. 
However, the knowledge on parton distribution functions (PDFs)
is still limited for many aspects of LHC physics and the discovery
potential is thus reduced.
In this proceeding, we show on one example that it is
possible to find some observables less sensitive to PDF uncertainties to probe
with a high efficency new physics beyond the standard model.

}

\FullConference{European Physical Society Europhysics Conference on High Energy Physics\\
                 July 16-22, 2009\\
                 Krakow, Poland}

\begin{document}

\section{Introduction}
At LHC, most measurements will be limited by systematic uncertainties. 
Experimental systematics can be reduced in ratios of quantities. 
In Ref. \cite{Boonekamp:2009yd}, 
we build appropriate ratios, for which the
sensitivity on theoretical uncertainties has been reduced compared to 
individual cross sections. The total error can then
be  reduced at a few percent level, 
showing that precision measurements at hadron colliders are possible.
We illustrate this recent work on one example.

\section{Drell-Yan production cross sections ratio}
With millions of $Z/\gamma^*$ produced with 1~fb$^{-1}$ of LHC data, 
the statistical error on the Drell-Yan production cross section is 
expected to be smaller than the percent. The limitation comes from 
systematics, among which a large error is due to PDF, around 6-8\%, 
even at high mass $\rm M>200~GeV/c^2$. The idea is to exploit the 
$Z/\gamma^*$ mass and rapidity spectrum and to make ratios of 
cross sections when the initial quarks have the same kinematics.\\

Let us consider a quark and anti-quark that produce a $Z$ boson with 
a rapidity $y$. The momentum fractions of these partons are
$x_1$$=M_Z/\sqrt{s}\cdot e^{-|y|}$ and $x_2=M_Z/\sqrt{s}\cdot e^{+|y|}$. 
But these momentum fractions can also be encountered in other 
$\gamma^*/Z$ processes. Symetric $q \bar{q}$ collisions with two partons 
carrying the momentum fraction $x_1$ can produce $\gamma^*/Z$ 
particles with the invariant mass $m=\sqrt{x_1 x_1 s}=M_Z\cdot e^{-|y|}$ 
and rapidity of 0. In the same way, symetric $q \bar{q}$ collisions with 
two partons carrying the momentum fraction $x_2$ can produce 
$\gamma^*/Z$ particles with the invariant mass $M=\sqrt{x_2 x_2 s}=M_Z\cdot e^{+|y|}$ 
and rapidity of 0. In other words, the same quark momenta have 
been found in three cross sections~: $\sigma_{Z/\gamma^*}(M_Z, y)$, 
$\sigma_{Z/\gamma^*}(m, y=0)$ and $\sigma_{Z/\gamma^*}(M, y=0)$ 
where $y=\ln M/M_Z$ and $m=M_Z^2/M$. Uncertainties on quark 
kinematics could be reduced in the following ratio, involving these 
three cross sections~:
$$
R(M)=\frac{\sigma(m, y=0)\cdot \sigma(M, y=0)}{\sigma^2(M_Z, y)}
$$
where $y=\ln M/M_Z$ and $m=M_Z^2/M$.\\

\begin{figure}[htbp]
  \begin{center}
    \includegraphics[width=.5\textwidth]{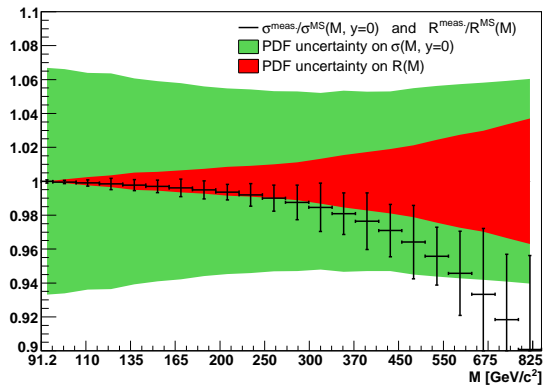}
  \end{center}
  \caption{\label{RatioSigma} Pseudo-measurement to Standard Model 
prediction ratios of $\sigma_{Z/\gamma^*}(M, y=0)$ or $R(M)$ with 
statistical error bars. The pseudo-measurements use $\rm 30~fb^{-1}$ 
of LHC data, and a 2~TeV SSM $Z'$ have been added in the simulation. 
The central values of  these two measurement are the same and the 
statistical uncertainties are very close, so only one set of error 
bars is shown. The uncertainty bands due to PDF on 
$\sigma_{Z/\gamma^*}(M, y=0)$ and $R(M)$ are also represented.}
\end{figure}

With only one quark flavour and with scale invariance, 
the PDF completely cancel and so their uncertainties. 
This is no longer valid in the real case but the prediction 
of $R(M)$ is still more precise than the high mass Drell-Yan 
cross section $\sigma(M, y=0)$. Fig.~\ref{RatioSigma} shows 
how these errors vary for different $Z/\gamma^*$ invariant masses. The PDF 
uncertainties can be reduced by more than a factor two, 
leading to a higher sensitivity to non-Standard Model processes.

An example of new physics sensitivity is shown on 
Fig.~\ref{RatioSigma}. Pseudo-measurement of $\sigma_{Z/\gamma^*}(M, y=0)$ 
and $R(M)$, including a 2~TeV SSM $Z'$ are compared to 
the Standard Model predictions. The statistical and PDF-induced 
uncertainties are also displayed. A  measurement of $R(M)$ 
shows a sensitivity of a 2~TeV SSM $Z'$ since $M>200~GeV/c^2$, 
while in a $\sigma_{Z/\gamma^*}(M, y=0)$ cross section analysis, 
no significant deviation is seen, except for $M>600~GeV/c^2$. 
Thus, it seems possible to explore a larger range of $Z'$ 
models, that may not be discovered by direct peak searches like 
non-resonant or wide $Z'$.

This method has other advantages. If a $Z'$ peak is observed, 
this ratio of cross sections can be used to measure the 
$\gamma^*/Z/Z'$ interference term at lower masses, in order to 
give additional constrains to the underlying $Z'$ model. 
Finally, this method can be applied to any $s$-channel processes 
like $W^\pm$ production. $W'$ searches or $s$-channel single-top 
cross sections can be normalized to $W^\pm\rightarrow l\nu$ 
to obtain more precise measurements.

\section{Conclusion}

Parton momentum density distributions are important ingredients in the
calculation of high energy hadron-hadron and lepton-hadron scattering
cross sections. In these calculations the cross sections are written
as a convolution of the parton densities and the elementary cross
sections for parton-parton or lepton-parton scattering. 
In Ref. \cite{Boonekamp:2009yd}, 
it is shown that the vision of the proton we have at present was
definitely improved with the recent data from HERA and Tevatron but
still suffers from large uncertainties at low or high $x$, with
significant impact on LHC plysics. No doubt that the understanding of
the proton structure will be further improved at LHC, and new
observables less sensitive to PDF uncertainties can be used to
disantangle in a better way the PDF effects from the ones due to new 
physics. 
In this proceeding, we have presented one particular observable,
defined as a ratio of Drell-Yan cross sections, for which the
sensitivity on theoretical uncertainties has been reduced compared to 
individual cross sections. Then, this ratio
shows a sensitivity of a 2~TeV SSM $Z'$ since $M>200~GeV/c^2$, 
where $M$ is the invariant mass of the DY system.
In contrast, for a $\sigma_{Z/\gamma^*}(M, y=0)$ cross section analysis, 
no significant deviation is seen, except for $M>600~GeV/c^2$.

\end{document}